# Ground-based Discovery of Cepheids and Miras in M101

David R. Alves[1] and Kem H. Cook[2]

IGPP Lawrence Livermore National Laboratory , Livermore, CA 94550

---

[1]Physics Department, University of California, Davis

[2]Visiting Astronomer Kitt Peak National Observatory

## ABSTRACT

We have identified 4 Cepheids and 5 Miras using KPNO 4m BVRI CCD images of an outer field in M101. The Cepheid and Mira periods range from 30 to 60 days and 350 to 800 days, respectively. We derive independent Cepheid and Mira distance moduli that agree within experimental uncertainties. We find a true distance modulus of 29.08 ± 0.13 mag.

*Subject headings:* stars: Cepheids, Miras — galaxies: distance

## 1. Introduction

M101 (NGC5457; RA 14 01.4, Dec 54 35; 1950) is the nearest supergiant Sc type galaxy. Its proximity, inclination, and high galactic latitude make it of prime importance for investigations of luminosity class I Sc spirals and their contents (Humphreys and Strom 1983). Observations of M101 have played a prominent role in efforts to measure the Hubble constant, $H_o$, which parameterizes the cosmological expansion rate and sets the extragalactic distance scale. In fact, discrepant distance estimates for M101 have long served as a symbol of the uncertainty in the extragalactic distance scale with those who support the short scale ($H_o = 90$ km s$^{-1}$ Mpc$^{-1}$) favoring an M101 distance modulus near 28.5 and supporters of the long scale ($H_o = 50$ km s$^{-1}$ Mpc$^{-1}$) arguing for moduli up to 29.5 (de Vaucouleurs 1993). Because of its small recession velocity and the potential confusion due to local gravitational perturbations, M101 cannot be used to directly measure $H_o$; however, the face-on inclination of M101 naturally lends itself to the search for secondary standard candles. It is largely for this reason, i.e. the calibration of secondary distance indicators capable of mapping the far reaches of the Hubble flow, that M101 has been the subject of extensive distance scale research. Several of the spiral members of the M101 cluster are also of interest because they can be used to calibrate the Tully-Fisher HI linewidth-luminosity relation. Advances in technology such as the Hubble Space Telescope have not relegated M101 to a position of historical curiosity. Secondary distance indicators will remain useful to astronomers as they continue to study distant galaxies and clusters of galaxies. The use of independent distance indicators in combination such as the Cepheid and Mira observations presented in this paper or the brightest supergiants and Cepheids in NGC4571 (Pierce 1992, Pierce et al. 1994) will help establish the scale of the Universe.

Using the KPNO 4m telescope we have found both Cepheid and Mira variables in a single moderately crowded spiral arm field. We have applied a Large Magellenic Cloud (LMC) Cepheid period-luminosity (PL) calibration and an LMC Mira PL calibration to derive independent relative distance moduli (M101-LMC) that agree within experimental uncertainties. Adopting our reddening corrected Cepheid value we estimate a modulus of 29.08 ± 0.13 mag (assuming an LMC modulus of 18.4 mag). In section II we describe the observations, the photometric reduction of the data, and the search for variable stars. In section III we discuss the Cepheid variables; deriving multi-wavelength PL calibrations for these stars and calculating a distance modulus for M101. In addition, we will discuss the implications of our reddening analysis for the Cepheids observed and investigate the dependence of the PL relations on the abundance gradient across the field. In section IV we discuss the identification of our Mira variables and employ a bolometric PL relation for an independent distance determination. In section V we discuss some of the implications of our M101 distance measurement such as the viability of ScI galaxies as standard rulers and the use of brightest red supergiants as standard candles. Section VI is our conclusion.

## 2. Observations and Data Reduction

CCD images of M101 and standard star fields were obtained over a four-year period with the KPNO 4-meter telescope. Using an 800×800 TI chip (TI2) at the prime focus yielded a pixel size of 0.3 arcsec and a full image area covering $4'\times4'$. The Mould $BVRI$ filter set was used. These filters cover a range in wavelength useful for fitting an analytical galactic extinction curve (Cardelli et al. 1989) to determine reddening. The $R$ band was selected as the primary variable search wavelength because of the improved seeing in the red and its optimal match to the sensitivity of the CCD. We believe that the advantage of using the most sensitive filter/instrument combination compensates for the smaller amplitude variation of Cepheids in $R$ than in $B$ or $V$.

On twelve nights standard stars in M92, NGC2419, NGC4147, NGC7790, and several

Landolt fields were observed (Landolt 1992, Davis 1984). The standards were observed with all filters, with short and long exposures, and at different airmasses during the night. The different filters were not exactly parfocal; but, in the interest of efficient observing some of the standard frames were taken out of focus. DAOPHOT II (Stetson 1987) was used to perform synthetic aperture photometry on the standard stars. Total instrumental magnitudes were calculated from aperture photometry with a radius of 10 pixels (3") and aperture corrections measured from several bright isolated stars in each field. In a final series of nightly calibration solutions, higher order extinction terms, including color-extinction terms, were found negligible. Each night's color coefficient was compared to the weighted average of all night's color coefficients and all were found to be consistent. Two nights (March 23, 1985 and April 4, 1986) had exceptionally well-constrained solutions and were chosen to place our program star photometry on the standard system. Figure 1 shows the residuals of our calibration solutions plotted against magnitude for the night of April 4, 1986.

Four separate fields in the spiral arms were selected for study. We report here on the full reduction of one field in an outer spiral arm. We have 17 epochs of R photometry covering a time span of 1176 days. In addition, we have 3 epochs of B, 2 epochs of V, and 4 epochs of I for this field. Most of the exposures are a cumulative 2700 or 3600 seconds; they are detailed in Table 1. Columns 1 - 3 of this table list the epoch, date, and number of days from epoch 1, respectively. Columns 4 - 7 contain the total exposure time in seconds for images obtained with the $B, V, R,$ and $I$ filters. Column 8 lists the R band photometric offset to epoch 11 and is described below. Column 9 lists the typical seeing in arcseconds for each R image. Separate 900 second exposures were preprocessed in the standard manner. The individual exposures were then offset and median combined alleviating the problem of cosmic ray hits in the final object images. Furthermore, bad columns and hot pixels in the CCD were typically offset in the shifted exposures and thus replaced with the median pixel value (typical of the sky) in the combined frames. The images in the best of these averaged frames have FWHM's of less than an arcsecond.

Photometry on the averaged M101 frames was performed with DAOPHOT II and ALLSTAR as prescribed by Stetson (1987). All of the stars were subtracted and the remaining image was smoothed; each pixel being replaced by the median value of a surrounding box approximately 20" across. The resulting smoothed "sky" frame was subtracted from the object frame. The median value of the "sky" frame was added to the sky subtracted object frame to maintain the appropriate signal level for noise calculations. In this manner, we eliminate the large scale background sky variations across the crowded spiral arms. A single epoch (March 23, 1985) with the best seeing and reaching the deepest was chosen to generate a master coordinate and photometry list for the field. An iterative process of finding stars, subtracting them with ALLSTAR, and then finding stars on the subtracted image was used to compile a list of 8000 stars. Coordinate transformations including linear, rotational, and scale terms were calculated and the master star list was transformed into each night's local coordinate system. ALLSTAR was not allowed to recenter these transformed coordinate lists. Empirical point spread functions (PSF's) were calculated with the same 3 isolated stars in each frame. This photometry gives R = 24 ± 0.10 mag (typical uncertainty for uncrowded stars) where this ALLSTAR error is derived from a goodness of profile fit and Poisson noise.

A set of potential internal standards (including our 3 PSF stars) was generated for this field by first compiling a list of isolated stars. Each isolated star was then paired with several others from the list and a magnitude difference for each pair was calculated in all 17 epochs. If a pair's relative magnitude differences showed a scatter about a mean value near that expected from the estimated photometric errors, but no more than 0.2 mag, both stars were considered non-varying. The typical standard deviation for each of these stars after being offset to a common instrumental magnitude system was 0.02 mag over all epochs. Photometric offsets relative to epoch 11 (the epoch chosen for its standard calibration solutions) with formal uncertainties of about 0.01 mag were calculated for the R frames from approximately 80 of the brightest internal standard stars. Stars with 14 of 17 possible epochs of photometry were tested for variability. A sigma-clipped reduced chi-square statistic was examined for the 5710 stars that met the minimum number of epochs criterium. The sigma-clipping removed each epoch of photometry from the reduced chi-square calculation that was over 10 standard deviations away from the weighted mean magnitude of all epochs. In this manner, a star's statistic was not skewed by a single epoch of very poor photometry. A simple cut left 503 candidate variable stars which were passed through a super-smoothing periodicity search routine (Riemann 1994). The 15 best fit periods were used to produce phased lightcurves for visual inspection. When inspecting the phased lightcurves, the possibility of confusion by an alias was carefully considered. A histogram of the the top 15 periods for all stars showed prominent peaks at common alias values. Our ground-based observation schedule produced these aliases because of the necessity of observing during dark/grey time. The morphology of the phased lightcurves was used to select proper periods; in particular, the amplitude variation and asymmetry (fast rise) of the curves were used to discriminate potential phased Cepheids and Miras. The most promising variable lightcurves were re-phased by hand where the period was varied in small increments to estimate the uncertainty. We find uncertainties ranging from 0.1 days for the short period 30 to 60 day Cepheids up to 2 days for the 350 to 750 day Miras.

In order to determine the stellar type of a phased variable, the color information was examined. We chose to consider flux-weighted mean colors for each variable. To calculate this, all 17 $R$ frames were offset to a common coordinate system and summed. 2 $B$, 2 $V$, and 4 $I$ frames were also summed. Note that the flux weighted colors derived from these small numbers of epochs in $B, V$, and $I$ are typically dominated by a single epoch with a long exposure. We present a $V, (V - I)$ color magnitude diagram for this field in Figure 2. The characteristic blue and red plumes are visible. The 503 variable candidates and several of our phased Cepheids and Miras are marked. A finding chart for the phased Cepheids and Miras is in Figure 3. These stars will be discussed in more detail in the following sections.

## 3. Cepheid Variables

Based on the lightcurve shape, periods, and colors we have identified 4 Cepheids. Two of the Cepheids (v1 and v2) were first found with 11 epochs of $R$ photometry (Cook et al. 1986). Using all 17 epochs we find the same periods as originally reported for these two stars. The Cepheids are described in Table 2. Column 1 identifies each variable, Column 2 lists the flux-weighted mean R magnitude, Columns 3 - 5 list the $(B - V)$, $(V - R)$, and $(R - I)$ colors, and Column 6 lists the period. Magnitude and color uncertainties quoted here are a combination of the error introduced by the standard calibration (0.02 mag), the error of the object frame aperture correction (ranging from 0.01 mag in $R$ to 0.08 mag in $V$), the error in zero pointing the instrumental magnitudes (0.02 mag, see section 2), and the ALLSTAR instrumental magnitude error. The phased lightcurves for these variables are in Figure 4.

To determine the distance to M101 from our Cepheids we must adopt a local calibration relating the period of pulsation to the absolute magnitude. We use multi-wavelength PL calibrations derived from LMC Cepheids (Madore and Freedman 1991). A distance modulus for M101 calculated from these calibrations will be relative to an LMC modulus; we adopt $18.4 \pm 0.10$ mag throughout this paper (de Vaucouleurs 1993). Using the slopes of the LMC lines, we fit the best relative LMC-M101 distance modulus to our M101 Cepheid period-luminosity data; this is shown in Figure 5. One might be concerned about a selection effect that appears to flatten the slope from the short period end by sampling the brightest Cepheids in the instability strip. The R band PL relation is marginally suggestive of such an effect. However, the relatively uniform distribution of data about the best fit lines in the other bands and the consistent moduli derived in each band help assure us that such a sampling effect is insignificant.

The derived moduli plotted against inverse wavelength are shown in Figure 6. Fitting the Cardelli galactic extinction curve to this plot allows us to determine the reddening affecting the Cepheids and use all the color information to simultaneously constrain the modulus. The line shown in Figure 6 is the weighted mean of the 4 moduli. The Cardelli galactic extinction curve extrapolated to long wavelengths gives a modulus within 0.02 mag of the weighted mean modulus; this line is not shown. We conclude that our data is consistent with zero reddening for these stars relative to the LMC calibrating Cepheids. We find this reasonable given the relative isolation of the Cepheids we have found and the observational bias for discovering unreddened variables (Cepheids in crowded regions are more likely to be affected by reddening). Some of the stars in this field are clearly reddened as is evident by the width of the blue plume in Figure 2 or a casual inspection of the finding chart in Figure 3. Observations at other wavelengths such as the J band could provide further information on the reddening. Without additional observations, there is no stronger contraint that can be placed on the reddening of these Cepheids than the analysis presented here utilizing data obtained at 4 different wavelengths. In addition, we assume zero galactic foreground reddening in the direction of M101 (galactic latitudes b > 50) (Sandage and Tammann 1974). This is consistent with the $HII$ region analysis of Zaritsky (Zaritsky 1990). Our derived modulus of 29.08 ± 0.13 mag is thus the true reddening corrected modulus. The uncertainty quoted has two sources: the uncertainty of the LMC modulus and the dispersion of the PL relations. Each contribute approximately equally.

In order to investigate abundance affects on the Cepheid PL relations we have located 6 $HII$ regions in our field from the Zaritsky study. These are split evenly with 3 in the northern half of the field and 3 in the south. The nucleus of M101 is to the S-SE. Excitation values are $\log([O_{III}]/H\beta)$ and have uncertainties of 0.1 dex. These can be converted into oxygen abundances, [O/H], via the empirical method which assumes a statistical relationship between the effective temperatures of the hottest stars in $HII$ regions and [O/H] (Edmunds and Pagel 1984). Assuming a solar (Fe/O) ratio we calculate a marginal metallicity gradient with values of -0.20 and -0.36 dex for the north and south portions of our field, respectively. Treating our northern and southern Cepheids separately we calculate a $BVRI$ modulus difference (north-south) of 0.03 $\pm$ 0.10 mag. The uncertainty in this relative modulus value is from the dispersion of the PL relations only. Thus we see no dependence of the PL relations on the small metallicity gradient across the field. Gould (1994) finds a metallicity dependence of the Cepheid PL relations in an analysis of $BVRI$ data on M31 and LMC Cepheids. He presents a modulus correction that depends on the metallicity of a field containing Cepheids relative to the mean LMC metallicity of -0.30 dex. This modulus metallicity correction predicts a north-south modulus difference of 0.10 $\pm$ 0.02 mag consistent with our results. Identification of Cepheids in our other M101 fields will provide a more significant test.

## 4. Mira Variables

Several of the long period variables (LPVs) we have found are likely Miras. We have identified approximately 50 red variables with amplitude variations in R greater than 1 magnitude. We have found periods for 4 of these stars and one red LPV with an amplitude variation less than 1 mag. This star, v7, has the longest period of our variables, with P = 758 days. It is possible that the observed small amplitude variation for v7 is attributable to the limited phase coverage; we may be missing the phase points of extreme of amplitude variation. We consider this a probable Mira. The phased lightcurves for the Miras are in Figure 7. The Miras are described in Table II.

Bolometric corrections were calculated for 4 of our 5 Miras from their $(V - I)$ colors and an empirical relation calibrated with oxygen-rich galactic, LMC, and SMC Miras (Bessel and Wood 1984).

$$BC_I = 0.3 + 0.38(V - I) - 0.14(V - I)^2$$

One variable, v8, has no measured $(V - I)$ color. We note the tendency for the longest period LPVs in the LMC (P > 450 days) to be oxygen-rich as opposed to carbon-rich (Hughes and Wood 1990). Since Miras exhibit large color changes with phase, our flux weighted colors (dominated by long exposures from different epochs) were not adequate for the bolometric corrections. First, we calculated the ratios of Mira amplitude variations in different filter bands to be $V/R = 1.9$ and $R/I = 1.5$ from $UBVRI$ observations of galactic Miras (Eggen 1969). The observed offset for each Mira from its epoch 3 R magnitude to the flux weighted mean R magnitude was scaled according to the calculated amplitude ratios and used to correct the epoch 3 $V$ and $I$ magnitudes to "pseudo-flux weighted mean" or "single epoch corrected" magnitudes. These single epoch corrected $V$ and $I$ magnitudes were used to derive $(V - I)$ colors for our 4 Miras. The bolometric corrections were fairly insensitive to the differences between $(V - I)$ colors calculated in this manner − corrected from epoch 11, adopted without corrections from epoch 3 or 11, or adopted from the flux weighted mean colors as described in section 2. The adopted $I$ magnitudes differed by approximately 0.1 to 0.4 mags if the naively calculated (4 epoch) flux weighted mean magnitudes were used instead of the single epoch corrected magnitudes.

We then calculated a bolometric PL relation from 24 (P > 400 day) LMC supergiant

Miras (Wood et al. 1983).

$$M_{bol} = -0.0025(P) + 12.266$$

Assuming that our 4 Miras are supergiant types, we calculate a relative M101-LMC modulus of 10.65 mag corresponding to an M101 modulus of 29.05 ± 0.15 mag. At this distance, all of our Miras have bolometric magnitudes above the maximum luminosity for asymptotic giant branch members, $M_{bol}$ = -7.1 (Wood et al. 1983); consistent with our supergiant assumption. The uncertainty quoted here is from the LMC modulus and an estimated dispersion about the 24 Mira LMC PL relation of approximately 0.10 mag. If we assume that our Miras are not supergiants but are instead long period members of the asymptotic giant branch we can use the bolometric LMC calibration (Hughes and Wood 1990) for AGB LPVs (P > 450 days)

$$M_{bol} = -3.22 - 7.76(logP - 2.4)$$

to calculate a relative M101-LMC modulus of 9.35 mag and an M101 modulus of 27.75 mag. In this case, all 4 Miras are consistent with being AGB members but the derived modulus is nearly a magnitude closer than other recent short scale distance determinations (de Vaucouleurs 1993). It is unlikely that these Miras are AGB members.

Near-infrared observations of 6 supergiant Miras in M33 and a K-band calibration with the same 24 LMC supergiants discussed above have been used by Kinman, Mould, and Wood to derive a relative LMC-M33 modulus of 6.14 mag (Kinman et al. 1987). We use the K magnitudes and a $(J - K)$ bolometric correction (Bessel and Wood 1984) to test our bolometric LMC calibration.

$$BC_K = 0.72 + 2.65(J - K) - 0.67(J - K)^2$$

We calculate a relative LMC-M33 modulus of 6.07 mag, consistent with the K-band measurement. We then derive a bolometric PL relation from these 6 M33 supergiant Miras

$$M_{bol} = -0.0024(P) + 18.279$$

and calculate a relative M33-M101 modulus of 4.60 mag from our 4 M101 supergiant Miras. Adopting an M33 modulus of 24.5 mag (de Vaucouleurs 1993) gives us an M101 distance modulus of 29.10 ± 0.15 mag. We note that some of the estimates of the M33 modulus are independent of the assumed LMC distance and in this respect the LMC-M101 and M33-M101 relative moduli are independent. The LMC Miras, M33 Miras, and M101 Miras are shown in Figure 8 where bolometric magnitude vs period (in days) is plotted. Also shown is the best LMC supergiant Mira PL line and this line offset to M33 and M101.

## 5. Implications of M101 Distance

The apparent correlation between absolute luminosity and the angular diameter of ScI galaxies is a distance indicator dependent on M101 for calibration (Sandage 1993). These easily recognizable high luminosity objects could be used as standard rulers in very distant clusters of galaxies. Sandage adopts an M101 distance modulus similar to our value and derives a low Hubble constant in a comparison with field spirals. With new Cepheid distances to the Virgo Cluster and M101 a direct test of this proposed standard ruler is now possible. We use M100 (NGC4321) in our comparison because it has the largest angular diameter of all Sc type spirals in Virgo and a recent Cepheid distance estimate of 17.1 Mpc (Freedman et al. 1994). We adopt (Tully 1988) the inclination corrected isophotal diameters of 6.0 and 23.8 arcmin for M100 and M101, respectively. The 10% uncertainties in the distance dominate the following analysis. Using our M101 Cepheid distance of 6.5 Mpc we calculate linear diameters in units of kpc of 29.8 ± 3.0 and 45.0 ± 4.5 for M100 and M101, respectively. M101 appears to be larger than M100.

Controversy concerning the calibration of the brightest red supergiants as standard candles has revolved around M101. Adopting an M101 distance modulus of 29.2 (Humphreys et al. 1986) the bolometric magnitudes for the brightest red supergiants observed imply masses violating stability limits in stellar structure theory. An even greater distance modulus would cause more discomfort. One would adopt at most a distance modulus of 28.4 to make the observations consistent with the upper luminosity boundary for M supergiants. Our distance modulus of 29.08 derived from Cepheids and Miras suggests an inconsistency between theory and observation worthy of further investigation. The apparent success of the brightest supergiant method for estimating a distance to NGC4571 (Pierce 1992) in agreement with the recent Cepheid distance is a further complication. It is possible that the brightest red supergiants scale with host galaxy luminosity (Sandage 1983).

## 6. Conclusions

We have found 4 Cepheids in one field in M101 using ground-based observations. These Cepheids yield a reddening corrected M101 distance modulus of 29.08 ± 0.13 mag. We have also found 5 Miras; 4 of which appear to be supergiants based on their bolometric magnitudes. We have applied an LMC supergiant bolometric PL calibration to derive an M101 distance modulus of 29.05 ± 0.15 mag and an M33 supergiant bolometric PL calibration to derive an M101 distance modulus of 29.10 ± 0.15 mag. The relative LMC-M101 distance moduli calculated from the Cepheids and Miras are independent. These moduli clearly support the long distance to M101 advocated by Sandage (Sandage and Tammann 1974).

Reduction of observations in 3 other fields of M101's spiral arms are in progress. The discovery of Cepheids in these fields will allow us to test the sensitivity of the Cepheid PL relations to the metallicity gradient across the face of M101. $J, H$, and $K$ near-infrared observations of our M101 Miras are planned to distinguish oxygen-rich from carbon-rich Miras and improve our bolometric corrections (Hughes and Wood 1990).

We thank Associated Western Universities for their support of David Alves' thesis

related research at a DOE facility. Work at LLNL is supported by the DOE under contract W7405-ENG-48. This work began as a collaboration between Marc Aaronson, Garth Illingworth, and one of us (Kem Cook). We acknowledge the strong foundation they provided and thank them. We also fondly remember Marc Aaronson and wish he could see the confirmation of this work by the HST Key Project he began.

– 16 –

---

- 18 -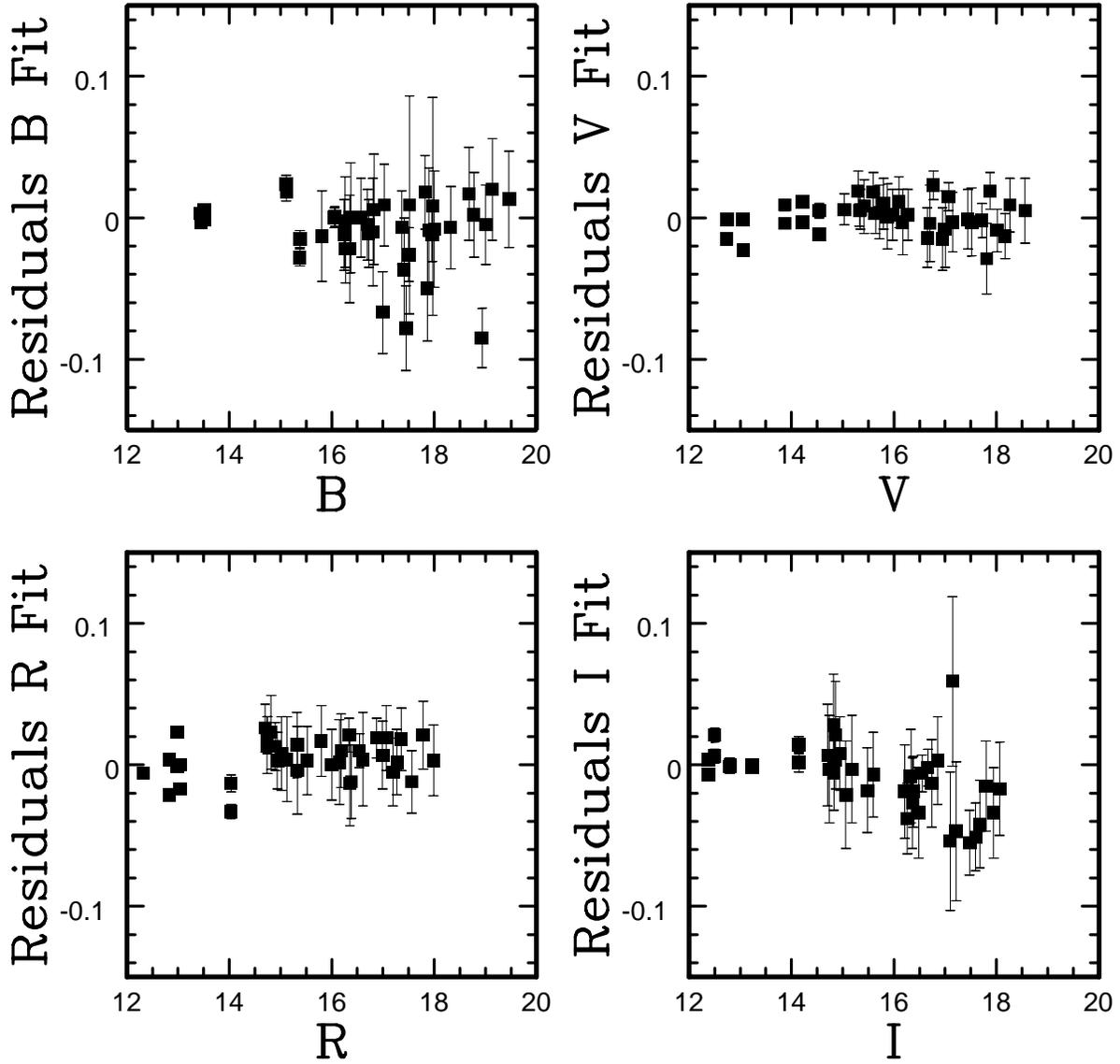

Fig. 1.— Residuals from photometric calibration solutions for April 4, 1986. 41 standard stars from M92, NGC4147, and Landolt-98 are used in the fits. Error bars on some of the brighter standards are smaller than the symbol used here.

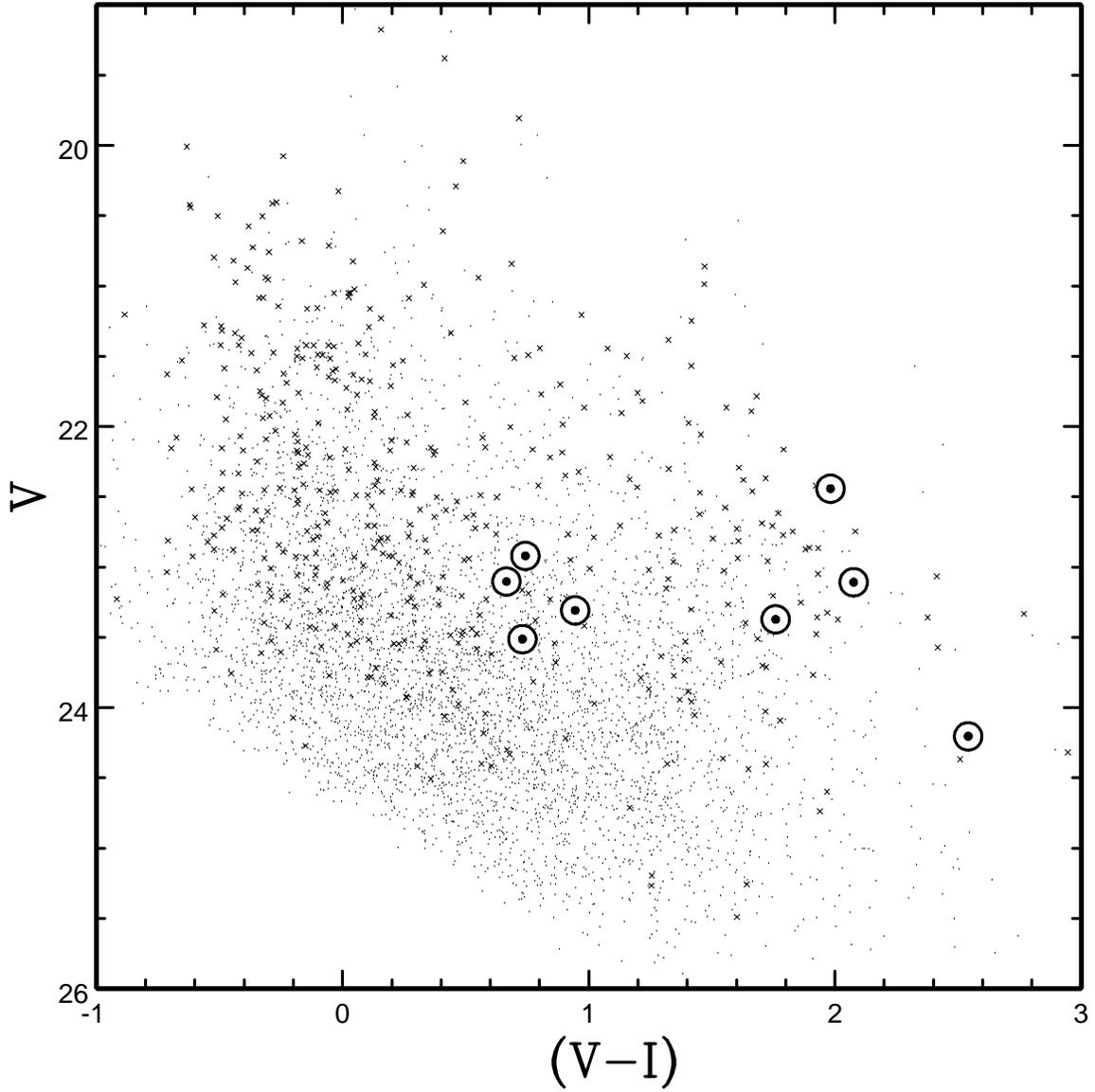

Fig. 2.— V,(V-I) color-magnitude diagram for M101 spiral arm field. Variable star candidates are denoted with a small $x$, phased Cepheids and Miras with a bullseye.

Fig. 3.— Finding chart for Cepheids and Miras in M101. Can be retrieved via anonymous ftp at igpp.llnl.gov, cd /pub/alves, get file m101_fig3.ps

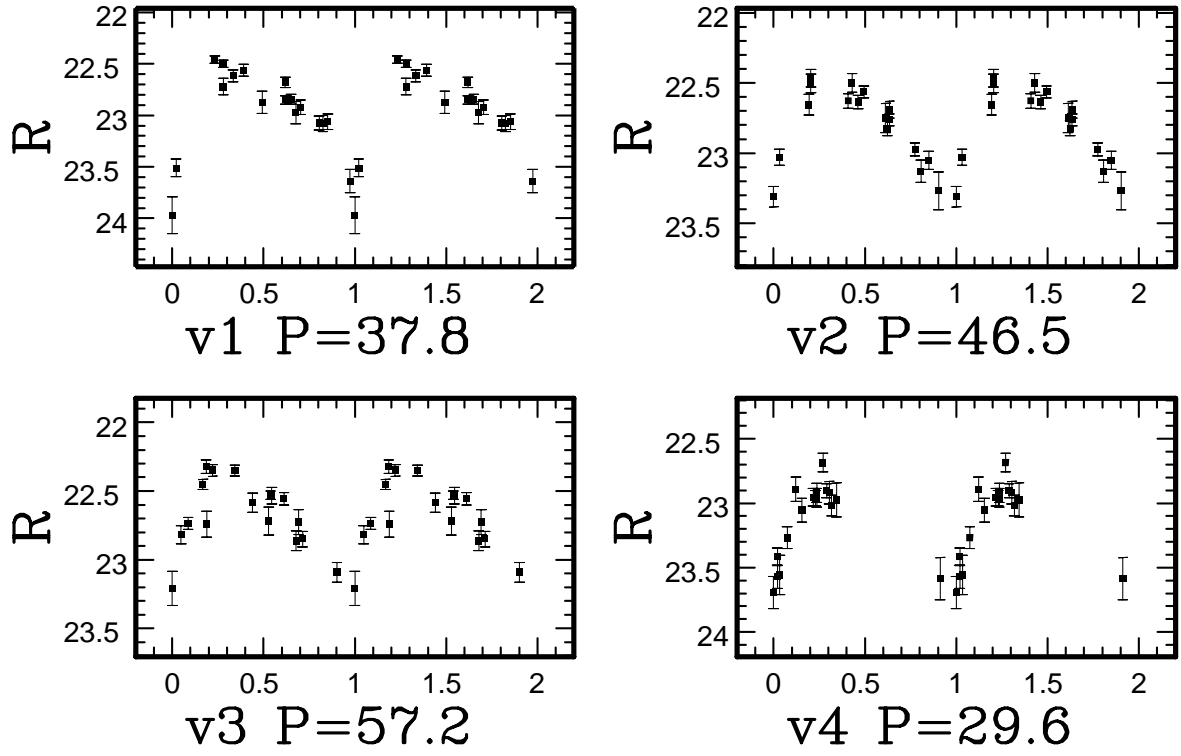

Fig. 4.— Phased lightcurves, magnitude vs phase, for 4 Cepheids. The faintest epoch of photometry has been set to phase = 0, each point has been plotted twice. Note the magnitude scale has been adjusted for each variable.

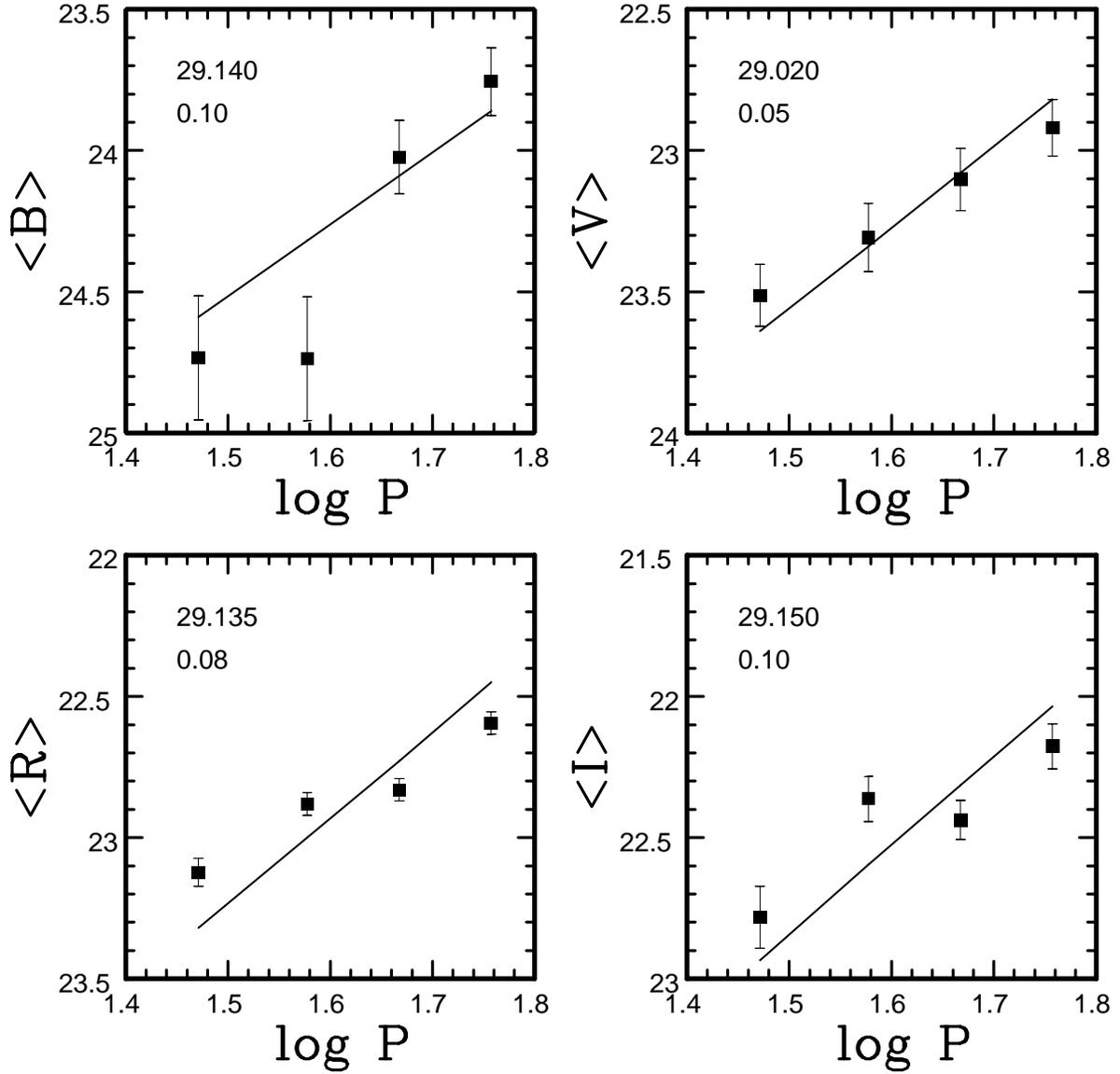

Fig. 5.— Cepheid Period-Luminosity relations from LMC fit to 4 M101 Cepheids in filter bands $B, V, R$, and $I$. The M101 moduli and uncertainty derived in each band are noted. Each plot is flux weighted mean magnitude vs log of period (days).

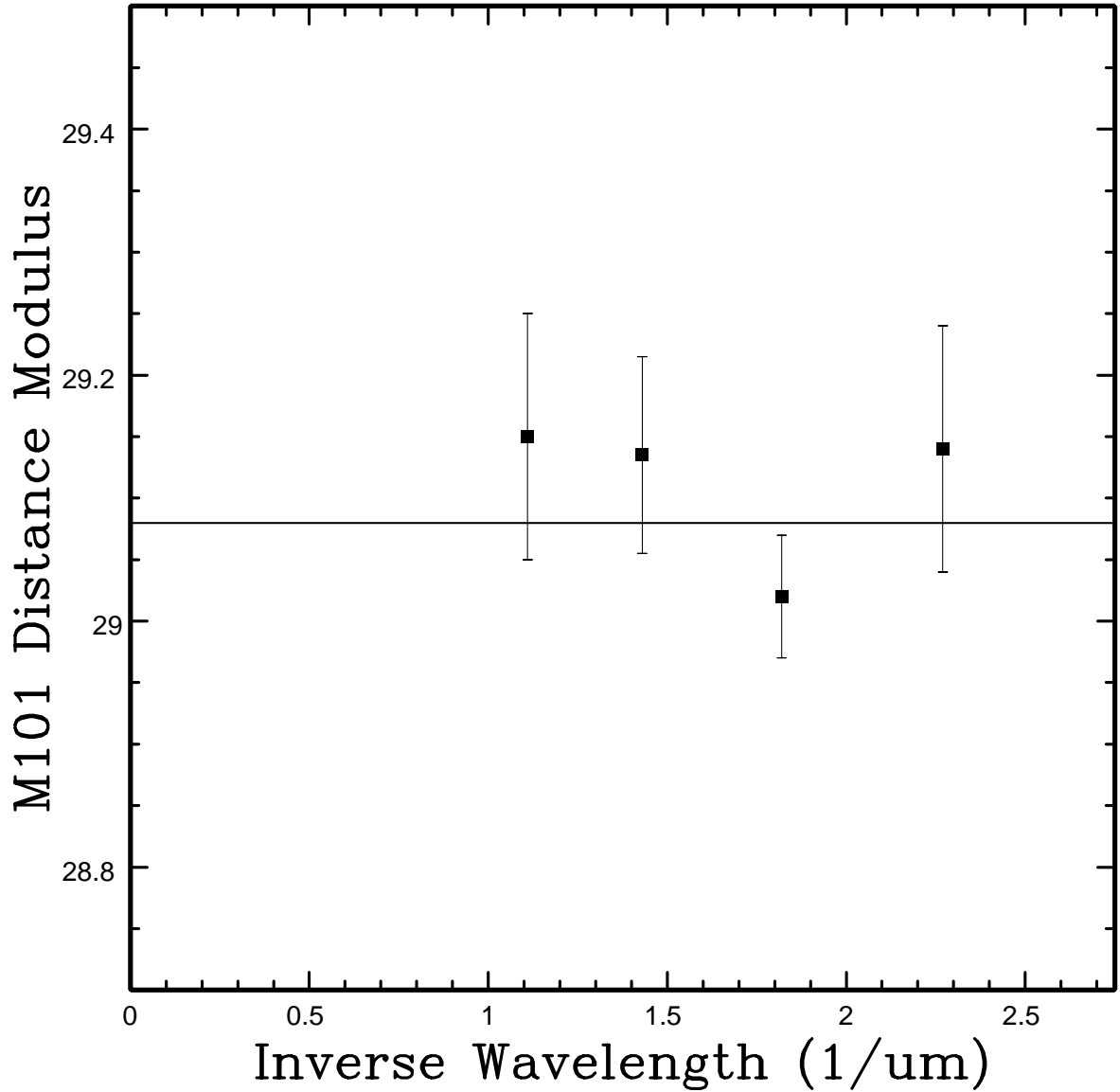

Fig. 6.— Derived M101 distance moduli plotted against inverse wavelength of filter band ($\mu m^{-1}$). The moduli are consistent with zero reddening for the M101 Cepheids relative to the LMC Cepheids. The fitted line is a weighted mean modulus. The galactic extinction law fit is consistent with zero reddening (a flat line).

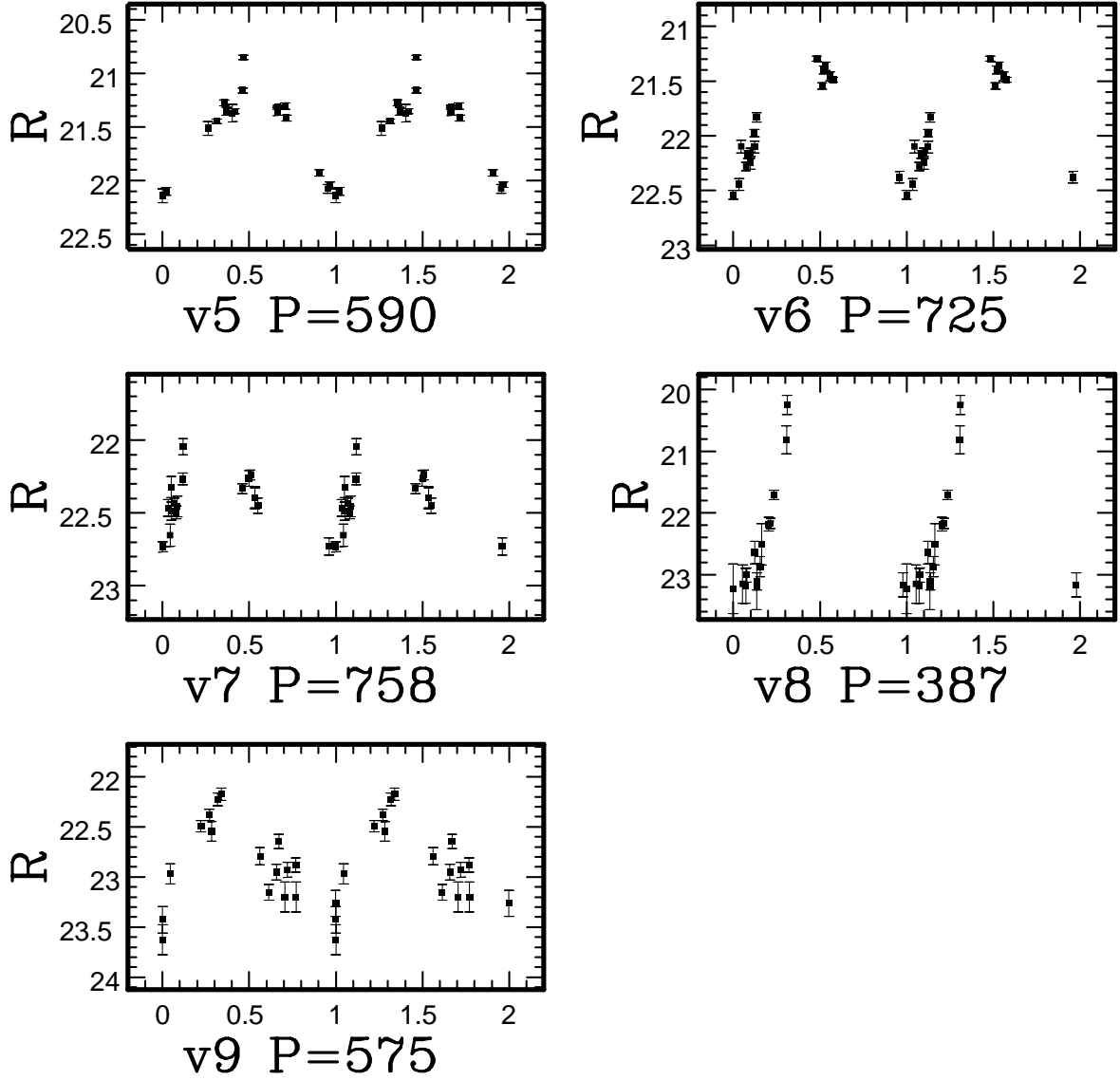

Fig. 7.— Phased lightcurves, magnitude vs phase, for 5 Miras. The faintest epoch of photometry has been set to phase = 0, each point has been plotted twice. Note the magnitude scale has been adjusted for each variable.

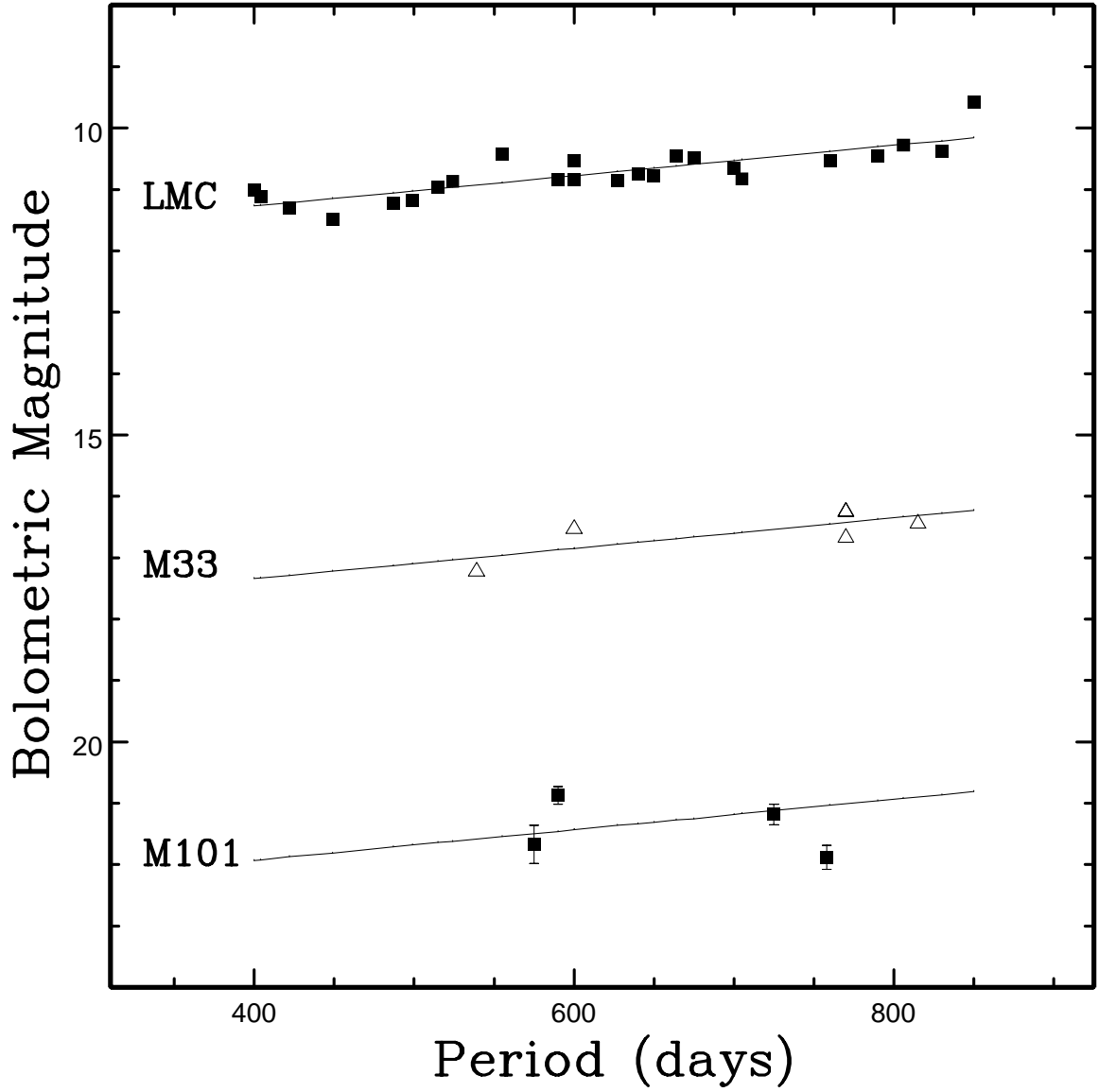

Fig. 8.— Bolometric magnitude vs. period (days) for 24 LMC supergiant Miras, 6 M33 supergiant Miras, and 4 M101 supergiant Miras.

TABLE 1. Summary of Observations

| epoch | date | day | B | V | R | I | offset | seeing |
|---|---|---|---|---|---|---|---|---|
| 1 | April 3, 1984 | 0 | — | — | 3600 | — | -0.353 | 1.3 |
| 2 | February 20, 1985 | 323 | — | — | 2700 | — | -0.016 | 1.0 |
| 3 | March 22, 1985 | 353 | — | 1800 | 2700 | 900 | 0.090 | 0.9 |
| 4 | April 16, 1985 | 378 | — | — | 3600 | — | -0.381 | 1.3 |
| 5 | April 24, 1985 | 386 | — | — | 2700 | — | -0.710 | 1.5 |
| 6 | May 14, 1985 | 406 | — | — | 3600 | — | -0.083 | 1.1 |
| 7 | May 22, 1985 | 414 | — | — | 3600 | — | -0.275 | 1.2 |
| 8 | June 19, 1985 | 442 | — | — | 2700 | — | -0.228 | 1.2 |
| 9 | June 20, 1985 | 443 | — | — | 2111 | — | -0.614 | 1.5 |
| 10 | March 6, 1986 | 702 | — | — | 3600 | — | -0.150 | 1.2 |
| 11 | April 4, 1986 | 731 | 400 | 300 | 2700 | 300 | 0.000 | 1.0 |
| 12 | April 11, 1986 | 738 | 3600 | — | 3600 | — | -0.494 | 0.9 |
| 13 | May 1, 1986 | 758 | — | — | 6300 | — | -0.746 | 1.2 |
| 14 | May 13, 1986 | 770 | — | — | 2700 | — | -0.225 | 1.2 |
| 15 | May 27, 1987 | 1149 | 4500 | — | 2700 | — | -0.693 | 1.5 |
| 16 | May 28, 1987 | 1150 | — | — | 2700 | 3600 | -0.296 | 1.1 |
| 17 | June 23, 1987 | 1176 | — | — | 3400 | 3600 | -0.667 | 1.4 |

Table 2. Phased Variable Stars

| id | $R$ | $B-V$ | $V-R$ | $R-I$ | Period |
|---|---|---|---|---|---|
| Cepheids | | | | | |
| v1 | 22.88 | 1.43 | 0.43 | 0.52 | 37.8 |
|  | 0.04 | 0.25 | 0.13 | 0.09 | |
| v2 | 22.83 | 0.92 | 0.27 | 0.39 | 46.5 |
|  | 0.04 | 0.17 | 0.12 | 0.08 | |
| v3 | 22.59 | 0.84 | 0.33 | 0.42 | 57.2 |
|  | 0.04 | 0.16 | 0.11 | 0.09 | |
| v4 | 23.12 | 1.22 | 0.39 | 0.34 | 29.6 |
|  | 0.05 | 0.25 | 0.08 | 0.12 | |
| Miras | | | | | |
| v5[s] | 21.53 | 1.32 | 0.91 | 1.07 | 590 |
|  | 0.03 | 0.19 | 0.10 | 0.05 | |
| v6[s] | 21.98 | 1.25 | 1.13 | 0.95 | 725 |
|  | 0.03 | 0.24 | 0.11 | 0.06 | |
| v7[s] | 22.42 | 1.16 | 0.95 | 0.81 | 758 |
|  | 0.04 | 0.30 | 0.11 | 0.06 | |
| v8 | 23.64 | — | — | 1.56 | 387 |
|  | 0.05 | — | — | 0.12 | |
| v9[s] | 23.24 | 0.85 | 0.97 | 1.57 | 575 |
|  | 0.07 | 0.33 | 0.17 | 0.10 | |

[s]supergiant

2